\documentclass[sigconf]{acmart}
\copyrightyear{2023}
\acmYear{2023}
\setcopyright{rightsretained}
\acmConference[ASSETS '23]{The 25th International ACM SIGACCESS Conference on Computers and Accessibility}{October 22-25, 2023}{New York, NY, USA}
\acmBooktitle{The 25th International ACM SIGACCESS Conference on Computers and Accessibility (ASSETS '23), October 22-25, 2023, New York, NY, USA}
\acmDOI{10.1145/3597638.3614505}
\acmISBN{979-8-4007-0220-4/23/10}
\title{Testaro}
\subtitle{Efficient Ensemble Testing for Web Accessibility}
\author{Jonathan Robert Pool}
\date{August 2023}
\acmSubmissionID{1001}

\begin{document}
\email{jonathan.pool@cvshealth.com}
\orcid{0000-0001-7864-5229}
\affiliation{%
  \institution{CVS Health}
  \streetaddress{1 CVS Drive}
  \city{Woonsocket}
  \state{Rhode Island}
  \country{USA}
  \postcode{02895}
}

\begin{abstract}

As automated web accessibility testing tools become enriched with new and improved tests, it can be impractical to leverage those advances. Each tool offers unique benefits, but effectively using multiple tools would require integrating them into a uniform testing and reporting scheme. Such integration is complex, because tools vary in what they try to detect, what they actually detect, and how they classify, describe, and report defects. Consequently, testers typically use only one tool.

\anon{Testaro}\cite{Testaro} is a novel open-source NPM package that checks compliance with about 650 rules defined by an ensemble of 8 tools: alfa, Axe, Equal Access, HTML CodeSniffer, Nu Html Checker, QualWeb, \anon{Testaro}, and WAVE.

Attendees at the demonstration will, within 5 minutes, create jobs for \anon{Testaro}, run them, and generate unified reports documenting more accessibility issues than any single tool can discover.

\end{abstract}

\ccsdesc[300]{Software and its engineering~Empirical software validation}
\ccsdesc[500]{Software and its engineering~Software testing and debugging}
\ccsdesc[500]{Human-centered computing~Accessibility design and evaluation methods}
\ccsdesc[500]{Human-centered computing~Accessibility technologies}
\ccsdesc[500]{Human-centered computing~Accessibility systems and tools}
\keywords{web accessibility, accessibility testing, test automation, test efficiency}
\received{23 June 2023}
\received[accepted]{6 August 2023}
\received[revised]{18 August 2023}

\maketitle

\section{Introduction}

Automated testing tools for web accessibility\cite{W3C1} have become abundant, cheap, and effective. Multiple open-source packages and free or inexpensive APIs make it practical to run automated linters during development, run automated tests after revisions, and repair the discovered defects before starting more expensive human testing.

Testing with an ensemble of tools, rather than only one, would be beneficial. Accessibility testing tools differ in the issues they try to detect and actually detect, and in how they classify, describe, and report issues.\cite{Englefield}\cite{Pool} Each tool discovers issues missed by others and casts doubt on other tools' findings.

The same diversity that makes it rewarding to test with an ensemble of tools also makes it difficult to do so. Integrating multiple tools into a uniform testing practice can be forbiddingly complex. Facing impractical integration costs, testers commonly trust a single tool for comprehensive web accessibility testing.

\anon{Testaro}\cite{Testaro}, an open-source NPM package, is designed to make accessibility testing with an ensemble of tools more efficient, by:

\begin{itemize}
\item installing all the tools
\item offering a uniform configuration language to select options for all the tools
\item running all the tools in a single job
\item extracting results from all the tools in a standard format
\item outputting the results in a single JSON job report
\end{itemize}

The demonstration will permit attendees to test URLs of their choice with \anon{Testaro} and inspect the results. Attendees will prepare jobs and post-process reports with a companion server package, \anon[ANONYMIZED2]{Testilo}\cite{Testilo}, compressing the test-running workflow into about two minutes of human effort and three minutes of automated testing.

\section{Related Work}

Previous research has generally found that accessibility testing tools substantially complement one another.\cite{Pool} Researchers have concluded that using only one testing tool is inadequate. Stanton\cite{Stanton}, for example, observed, ``no one tool gave me a complete list of issues found by the others'', and concluded, ``you really do have to test in multiple tools.''

\anon{Pool}\cite{Pool} obtained similar results testing 121 web pages with 9 tools. He found no tool redundant. Classifying the rules of those tools into 245 ``issues'', he reported that removing one tool from the set would cause at least 7 of the reported issues to be lost. Moreover, tools reporting the same issues sometimes reported very different counts of instances. For example, 8 of the tools discovered links without accessible names, but their instance counts ranged from 2 to 240.

An in-principle solution to the complexity of ensemble accessibility testing was proposed at IBM in 2005 by Englefield, Paddison, Tibbits, and Damani \cite{Englefield}: a single platform that would provide user interaction, result processing, and other shared services for tools, allowing tools to be nothing more than rule engines. That solution, however, has remained an ambitious idea. Tool makers have not cooperated to implement it, and the inventory of tools remains fragmented and incoherent.

Until now, no project integrating multiple accessibility testing tools under programmatic control with standardized reporting has been discovered. Pa11y\cite{Pa11y}, kayle\cite{kayle}, and AATT\cite{AATT} integrate 2 tools: Axe and HTML CodeSniffer. Although a11yTools\cite{Adam} integrates 5 tools and 13 single-issue tests, it runs only one tool or test at a time, and only under human control.

\section{Architecture}

\anon{Testaro} (in contrast with the Englefield \textit{et al.} proposal) does not depend on cooperation from tool makers. It integrates existing tools as they are.

\anon{Testaro} tests the way humans do. It launches web browsers, navigates to web pages, performs actions, checks whether the pages behave as expected, and notes the results. Hence, it runs on a Windows, MacOS, or Ubuntu workstation.

\anon{Testaro} is an NPM package that performs its own tests and those of 7 other tools, of which one is a remote service and the others are installed dependencies. The tools integrated by \anon{Testaro} are listed in Table~\ref{tab:pkgs}. Among them, they check compliance with about 650 rules. \anon{Testaro} uses Playwright\cite{Pw} to launch and control Chromium, Webkit, and Firefox browsers.

\begin{table}
  \caption{Tools integrated by \anon{Testaro}}
  \label{tab:pkgs}
  \begin{tabular}{lll}
    \toprule
    Code & Name & Creator\\
    \midrule
    alfa & alfa\cite{alfa} & Siteimprove\\
    axe & axe-core\cite{axe} & Deque\\
    htmlcs & HTML CodeSniffer\cite{htmlcs} & Squiz\\
    ibm & Equal Access\cite{ibm} & IBM\\
    nuVal & Nu Html Checker\cite{nuVal} & W3C\\
    qualWeb & QualWeb\cite{qualWeb} & Universidade da Lisboa\\
    \anon{testaro} & \anon{Testaro}\cite{Testaro} & \anon{Testaro}\\
    wave & WAVE\cite{wave} & WebAIM\\
    \bottomrule
  \end{tabular}
\end{table}

\section{Process}

A \textit{job} is an object giving information and instructions to \anon{Testaro}. The core of a job is its \textit{acts}, an array of instructions to be executed. Version 18.0.0 of \anon{Testaro} defines 19 act types, which include actions on a page, navigations among pages, and tool executions.

When an act tells \anon{Testaro} to execute one of the 9 tools, the act can specify which rules of that tool the tool should test for, which of the 3 browser types the tool should use, how granular the output should be, and other options. Here is an example of an act, telling \anon{Testaro} to make the \texttt{alfa} tool perform tests for two of its rules:

\begin{verbatim}
{
  type: 'test',
  which: 'alfa',
  what: 'Siteimprove alfa tool',
  rules: ['r25', 'r71']
}
\end{verbatim}

As it performs a job, \anon{Testaro} adds results to the acts. At the end of the job, \anon{Testaro} adds whole-job data to the job and returns this elaborated job as a \textit{report}.

\section{Efficiencies}

\anon{Testaro} is designed to streamline tool installation and configuration. It installs all the tools and provides a uniform configuration interface. The options made available by all the tools are documented in one location and selected in the job file with a uniform syntax.

\anon{Testaro} simplifies the task of executing multiple tools. A single job file tells \anon{Testaro} which tools to launch in which order, and \anon{Testaro} runs them all. A job that includes all the tests of all the tools typically takes about 3 minutes. If that were not fast enough, execution could be further accelerated with job partitioning: installing \anon{Testaro} on multiple workstations, having them perform complementary jobs in parallel, and combining their reports.

An instance of \anon{Testaro} can be configured as an on-call agent. It polls a server for jobs. When the server replies by sending a job, \anon{Testaro} performs it and sends the report to the server.

Finally, \anon{Testaro} is designed to make the utilization of tool reports more efficient. For this purpose, \anon{Testaro} translates the most common elements of native tool reports into standard results. Fully documented in the \texttt{README.md} file, the standard results uniformly present each tool's reports of violations of its rules, including what rule was violated, how serious the estimated impact is (on a 0-to-3 ordinal scale), what HTML element was involved, where on the page it appeared, and an excerpt from the HTML code.

Here is an example of an entry from a standard result:

\begin{verbatim}
{
  totals: [23, 11, 6, 8],
  instances: [
    {
      ruleID: 'image-no-alt',
      what: img element has no text alternative,
      count: 1,
      ordinalSeverity: 3,
      tagName: 'IMG',
      id: 'ocean-beach-sunset',
      location: {
        doc: 'dom',
        type: 'xpath',
        spec: '/html/body/div[4]/p[2]/img[1]'
      }
      excerpt: <img src="images/obSunset.jpg">
    },
    ...
  ]
}
\end{verbatim}

In this example, a tool reported 23 instances of rule violations at severity 0, 11 at severity 1, etc. The first reported instance was an \texttt{img} element that violated a rule named \texttt{image-no-alt}.

\begin{table*}
  \caption{Targets}
  \label{tab:targets}
  \begin{tabular}{lll}
    \toprule
    ID & Description & URL\\
    \midrule
      w3c & World Wide Web Consortium & https://www.w3.org/\\
      mozilla & Mozilla Foundation & https://foundation.mozilla.org/en\\
      wikFnd & Wikimedia Foundation & https://www.wikimedia.org/\\
      acm & Association for Computing Machinery & https://www.acm.org/\\
    \bottomrule
  \end{tabular}
\end{table*}

Given the diverse ontologies of the tools, any standardization reflects some judgment. An example is the \texttt{ordinalSeverity} property, which interprets and combines the tools' various classifications of severity, priority, and certainty. Users are free to rely on the standardization performed by \anon{Testaro} to simplify report consumption, but, if they want more control, they may extract data from original tool results, too.

\section{Customization}

Effective accessibility management requires checking conformity not only to industry standards such as the Web Content Accessibility Guidelines\cite{W3C2} (WCAG), but also to rules (brand standards, design systems, etc.) of one's own organization. In a multi-tool integrator, each tool is potentially a platform for the creation of custom rules, and the set of tools is extensible. Users can customize \anon{Testaro} by any of these methods:

\begin{itemize}
\item creating a tool and adding it as an installed dependency
\item creating a tool and adding it as a remote service
\item extending any of the tools, if it permits, by adding new rules to it
\end{itemize}

The \anon{Testaro} tool contains a template for the creation of custom rules. Existing Testaro rules are typically defined in 10 to 30 lines of code. A custom rule would likely require a similar amount of code.

\section{Job preparation}

For routinized use of \anon{Testaro}, job preparation can be partly automated. One package performing this function is \anon[ANONYMIZED2]{Testilo}\cite{Testilo}. The user can create files that answer the questions ``What tests do you want to run?'' and ``What \textit{targets} do you want to test?''. \anon[ANONYMIZED2]{Testilo} can convert those files to a job that \anon{Testaro} will execute. Table~\ref{tab:targets} gives an example of data that might be in a target file.

\section{Report Enhancement}

A JSON report from \anon{Testaro} narrows the gap between native tool reports and user-friendly reporting, but does not close that gap. A report contains standard results, but they are presented sequentially, tool by tool, and the result from each tool describes violations of that tool's rules, not of universally defined norms. Users will often want to:

\begin{itemize}
\item map the tool rules onto a set of tool-agnostic issues
\item gather the complaints of all tools about each issue into one place
\item aggregate the issue reports into a total accessibility score
\item export scores for use in dashboards or reports
\item summarize the JSON report in a developer- or manager-friendly HTML document
\item collect scores from reports on related targets into a comparative report
\end{itemize}

To perform such functions, users can create procedures and/or use \anon[ANONYMIZED2]{Testilo}. To interpret tool rules, \anon[ANONYMIZED2]{Testilo} offers a rule classifier that maps the approximately 650 tool rules onto about 260 \textit{issues}. For example, two tools have rules prohibiting broken same-page links. One is \texttt{AAA.2\_4\_1.G1,G123,G124.NoSuchID} from \texttt{htmlcs}, and the other is \texttt{link\_internal\_broken} from \texttt{wave}. \anon[ANONYMIZED2]{Testilo} maps both of these onto an \texttt{internalLinkBroken} issue and references WCAG Success Criterion 1.3.1 as the most relevant standard.

\section{Demonstration}

In the demonstration, a simple web service will ask each user for the URL of a web page to be tested. The service will use \anon{Testilo} to create a job for \anon{Testaro}. \anon{Testaro} will perform the job and convert all the tools' results into standard results. As the final step, \anon{Testilo} will convert the \anon{Testaro} report to a human-oriented web page.

\section{Future work}

Some engineers using \anon{Testaro} for accessibility testing have requested richer and more tailored issue reports, with better identification of instance locations, consolidation of duplicates, and resolution of tool disagreements.

Such improvements will require more work on instance identification. Some tools may locate instances by line number, others by XPath, others by CSS selector, others by bounding box, and others only with an HTML code excerpt. Further work could aim to determine when instances reported by various tools are the same and to supply images of, and links to, instances.

Empirical data from the use of \anon{Testaro} may facilitate rule sequencing, pruning, and deprecation; bug reports to tool makers; and the training of machine learners in the prediction of accessibility issues.
https://www.overleaf.com/project/64d0165e85f6f56474c88ea3
\anon{Testaro} welcomes contributions to improve functionality, reliability, and issue coverage.

\section{Conclusion}

As makers of testing tools innovate to narrow the gaps\cite{Nielsen}\cite{Matuzo} between formal and practical accessibility, tools continue to complement each other. Accessibility testing with an ensemble of tools is complex but valuable, and it can be made practical even without the cooperation of tool makers.

\begin{acks}

I acknowledge valuable editorial comments and research from \anon{Susan M. Colowick}.

My opinions expressed herein are my own views and do not necessarily reflect the views of \anon{CVS Health}, its affiliates, or any of my colleagues at \anon{CVS Health} or its affiliates.

\end{acks}

\bibliographystyle{ACM-Reference-Format}
\bibliography{main}

\end{document}